\documentclass[doublecol]{epl2}

\usepackage{amsmath}
\usepackage{amssymb}
\usepackage{graphicx}
\usepackage{dcolumn}
\usepackage{bm}
\usepackage{epsfig}
\usepackage[normalem]{ulem}

\title{Lieb's soliton-like excitations in harmonic trap}

\author{G.E. Astrakharchik\inst{1} \and L.P. Pitaevskii\inst{2,3}}

\institute{
  \inst{1} Departament de F\'{\i}sica i Enginyeria Nuclear, Campus
Nord B4-B5, Universitat Polit\`ecnica de Catalunya, E-08034 Barcelona, Spain \\
  \inst{2} INO-CNR BEC Center and Dipartimento~di~Fisica,
Universit\`{a} di Trento, I-38123 Povo, Trento, Italy \\
  \inst{3} P.L.~Kapitza Institute for Physical Problems RAS, Kosygina 2, 119334 Moscow, Russia.
}

\abstract{
We study the solitonic Lieb II branch of excitations in one-dimensional Bose-gas in homogeneous and trapped geometry. Using Bethe-\textit{ansatz} Lieb's equations we calculate the ``effective number of atoms'' and the ``effective mass'' of the excitation. The equations of motion of the excitation are defined by the ratio of these quantities. The frequency of oscillations of the excitation in a harmonic trap is calculated. It changes continuously from its ``soliton-like'' value $\omega_h/\sqrt{2}$ in the high density mean field regime to $\omega_h$ in the low density Tonks-Girardeau regime with $\omega_h$ the frequency of the harmonic trapping. Particular attention is paid to the effective mass of a soliton with velocity near the speed of sound.
}

\begin{document}

\maketitle

\section{Introduction} Recent development of experimental techniques has opened an exciting possibility to work with ultracold Bose gases in one-dimensional (1D) conditions, for example in a set of elongated optical traps (see Moritz et al.~\cite{Essl03}) and in magnetic traps created by solid state chips (Esteve et al.~\cite{Esteve06}). This development permitted to verify the theoretical predictions in highly controllable experiments. The theoretical investigation of one-dimensional bosons has been began by Marvin Giradeau in Ref.~\cite{Gir60}, where the case of an infinite repulsion was considered. This case is often called as ``Tonks-Giradeau'' (TG) limit, although Tonks considered 1D classical gas~\cite{Tonks36}. The next important step was made by Lieb and Liniger  in Refs.~\cite{LiebLiniger63,Lieb63} where they obtained an exact solution of the problem of ground state properties and energy spectrum of the one-dimensional Bose gas with the delta-functional repulsive interaction (Lieb-Liniger gas). Probably, the most surprising result of the paper~\cite{Lieb63} is the existence, besides the phonon-like branch of elementary excitations (Lieb I branch), which presence was natural to assume in analogy with 3D case, also of the second branch (Lieb II branch). This branch exists in a finite interval of the momenta $|p|/\rho  \leq \pi$ and its energy approaches zero, when the coupling constants tends to zero. In TG limit the spectrum coincides with that of an ideal Fermi gas and Lieb II branch corresponds to excitation of holes. The meaning of the second branch in the opposite (mean-field ``Bogolyubov'') limit of weakly interacting bosons was explained by Kulish, Manakov and Faddeev~\cite{Fadd76} (see also Ishikawa and Takayama~\cite{Ish80}). They have shown, that the energy-momentum dispersion relation for the second branch in this limit coincides with the relation, obtained by Tsuzuki \cite{Tsu71} for a soliton, described by the Gross-Pitaevskii equation (GPE). Recently Sato et al. have shown that the spatial profile of the order parameter, defined as a proper matrix element, also reproduces the GPE soliton profile~\cite{Sato12}. It was also shown by Kanamoto, Carr and Ueda that  states with non-zero angular momenta  of Lieb's Hamiltonian on a ring can be identified in the same limit as multisoliton solutions of GPE~\cite{Carr10}. Hence Lieb II branch of excitations in the intermediate regime is a result of a quite non-trivial crossover between a topological soliton and an excitation of fermion-like holes. Investigation of the properties of these unusual objects is, in our opinion, an interesting and important problem. In this paper we investigate dynamics of Lieb II branch of excitations in a gas confined to 1D harmonic trap. We assume that the size of the cloud is large in comparison with the healing length. Then one can safely use the Local Density Approximation (LDA) for the dynamics.

\section{Local Density Approximation} In the LDA, dynamics of an excitation is defined by its dispersion law in a uniform gas. The most convenient description of the dynamics is in terms of energy of the excitation $\varepsilon(V,\mu)$ expressed as a function of its velocity $V$ and the chemical potential $\mu$. For a smooth external potential $U(x)$, the LDA energy can be obtained by exchanging the chemical potential by its local value $\mu \rightarrow \mu-U(x)$. This means that in the course of the motion of an excitation in the presence of the external field, the energy $\varepsilon (V,\mu -U(x))$ must remain constant \cite{Shl99,KonPit04}. Differentiating with respect to time and taking into account that $dx/dt=V$
we obtain $1/V(\partial \varepsilon /\partial V)_{\mu }dV/dt-(\partial \varepsilon /\partial \mu )_{V}(\partial U/\partial x)=0$, or
\begin{equation}
m_{eff}\frac{dV}{dt}=-N_{s}\left( \frac{\partial U}{\partial x}\right)
\label{Newt}
\end{equation}
where the parameters, characterizing the excitations,
\begin{eqnarray}
\begin{array}{lll}
m_{eff} &=&\frac{1}{V}\left( \frac{\partial \varepsilon }{\partial V}\right)_\mu\\
N_s &=&-\left( \frac{\partial \varepsilon }{\partial \mu }\right) _{V}
\end{array}
\label{meffNs}
\end{eqnarray}
have, correspondingly, meaning of the effective mass $m_{eff}$ and the effective number of atoms $N_s$ in the excitation. For the excitations of the second branch in the Bose gas these quantities are negative, thus $|N_s|$ is the number of atoms expelled at creation of an excitation. These equations of motion of solitons in LDA approximation were derived in \cite{KonPit04} for the GPE solitons and in \cite{Scott11} for the general case.

The effective number of atoms in an excitation $N_s$ appears in a natural way in the equation for $dV/dt$. However, one should take into account, that it is not identical to the number of atoms $N_d$, introduced in \cite{Schecter12}. These quantities coincide in the Bogolyubov limit.

It is convenient to rewrite the equation (\ref{Newt}) as
\begin{eqnarray}
\label{Z}
Zm\frac{dV}{dt}=-\left( \frac{\partial U}{\partial x}\right),\quad
Z(V,\mu )=\frac{m_{eff}}{mN_s}
\end{eqnarray}
where $m$ is mass of an atom. Dimensionless ``mass renormalization'' function $Z$ is the only parameter, describing dynamics of the soliton in LDA.

Quantities $m_{eff}$ and $N_s$ can be easily calculated in the Bogolyubov regime using GPE. Here according to~\cite{Tsu71} the energy of a soliton is $\varepsilon (V,\mu )=2\hbar (\mu -mV^2)^{3/2}/(3cm^{1/2})$, where $2c$ is the one-dimensional coupling constant (see Eq.~(\ref{H}) below). Correspondingly
\begin{eqnarray}
\label{Eq:GP Ns}
\begin{array}{lll}
N_s&=&-\frac{\hbar }{cm^{1/2}}\left(\mu -mV^2\right)^{1/2}\\
m_{eff}&=&2mN_s
\end{array}
.
\end{eqnarray}
Thus in the Bogolyubov regime the effective mass of a soliton is twice the total mass $N_sm$ of the particles in it, so as far as dynamics are concerned, a GPE soliton moves in an arbitrary external field as a particle of mass $2m$~\cite{KonPit04}.

If the gas is trapped in a harmonic trap $U(x) = m\omega_h^2x^2/2$, the frequency of small oscillations can be found from the equation of motion~(\ref{Newt}) keeping the values of $N_s$ and $m_{eff}$ constant and equal to the ones at $V=0$ and in the center of the trap. The frequency of harmonic oscillations depends on soliton properties as
\begin{equation}
\Omega =\sqrt{\frac{mN_s}{m_{eff}}}\omega_h=\frac{1}{\sqrt{Z}}\omega_h\;.  \label{Omega}
\end{equation}
For the GPE soliton one has $Z=2$ and $\Omega =\omega_h/\sqrt{2}$. This result was first obtained in \cite{Ang00} by a different method and confirmed in experiments\cite{Seng08}.

In the opposite, TG limit the energy of an second-branch excitation can be presented as $\varepsilon(V,\mu )=\mu -mV^2/2$ and
\begin{eqnarray}
N_s =-1,\quad
m_{eff} = -m
\label{Eq:TG meff}
\end{eqnarray}
correspondingly to the ``hole-like'' nature of the excitation in this limit. In this case the frequency of oscillations is $\Omega = \omega_h$. In this paper we will calculate the characteristic parameters $m_{eff},N_{s},Z$ and the frequency $\Omega$ for intermediate strengths of the interaction.

It is worth noticing that in absence of external field the state with one soliton in the Lieb-Liniger model has an infinite lifetime. In the presence of trapping an excitation has a finite lifetime due to emission of phonons. This effect has been investigated in~\cite{Gang12} for the GPE solitons. The probability of the decay is small at small enough $\omega_h$. In the following we will not consider this effect.

\section{Lieb's equations}

In the Lieb-Liniger model the Hamiltonian is written as
\begin{equation}
H=\frac{\hbar^2}{2m}\sum_i\frac{d^2}{dx_i^2}+2c\sum\limits_{i<k}\delta(x_i-x_k) .  \label{H}
\end{equation}
In the original paper\cite{LiebLiniger63} authors used the system of units with $\hbar=1,m=1/2$. Calculation of the second branch of the spectrum of elementary excitations is reduced to the solution of a linear integral equation for the function $J(k,q)$
\begin{equation}
2\pi J(k,q)-2c\int\limits_{-K}^{K}\frac{J(r,q)dr}{c^2+(r-k)^2}
=\pi-2\tan^{-1}\left(\frac{q-k}{c}\right).
\label{Eq:J(k,q)}
\end{equation}
The limit of integration $K$ defines the one-dimensional density $\rho$ (and the value of the dimensionless parameter $\gamma=c/\rho$) indirectly, as an integral of the solution of an equation similar~(\ref{Eq:J(k,q)}), but without $q$-dependent term on the r.h.s. \cite{LiebLiniger63}. The dependence $K(\gamma)$ and the inverse one $\gamma(K)$ can be calculated following the methods of \cite{LiebLiniger63}. Once such relations are known, the sound velocity $u$ can be calculated according to $u=-2\gamma^2d(K/\gamma)/d\gamma$. We use matrix methods to solve Eq.~(\ref{Eq:J(k,q)}) and similar integral equations. To do so we discretize the integral which than is written as a $(r,q)$ matrix. The inverse matrix is calculated and is multiplied by the discrete representation of the r.h.s. of Eq.~(\ref{Eq:J(k,q)}).

The knowledge of $J(k,q)$ permits to calculate the dependence of the energy $\epsilon$ on the momentum $p$ in the parametric form (here $q$ is understood as a free parameter):
\begin{equation}
\begin{array}{lll}
\varepsilon &=&\mu-q^2+2\int\limits_{-K}^{K}J(k,q)k\;dk\\
p&=&-q+\int\limits_{-K}^{K}J(k,q)\;dk
\end{array}
.
\label{ep}
\end{equation}

The resulting energy $\varepsilon$ of excitations is a function of $\mu$ and $p$, instead of $\mu$ and $V$. It is possible to calculate $V(p,\mu)=(\partial \varepsilon/\partial p)_\mu$ from equations~(\ref{ep}). To calculate $N_s$ and $m_{eff}$ in these variables one can use the relations
\begin{eqnarray}
\begin{array}{lll}
m_{eff}&=&
\frac{1}{V}
\left(\frac{\partial\varepsilon}{\partial p}\right)_\mu
\left(\frac{\partial p}{\partial V}\right)_{\mu},\\
N_s&=&
-  \left(\frac{\partial\varepsilon}{\partial \mu}\right)_p
+V \left(\frac{\partial V}{\partial\mu}\right)_p
   \left(\frac{\partial p}{\partial V}\right)_{\mu}
\end{array}
.
\label{Ns}
\end{eqnarray}

The natural parameters of Lieb's equations (\ref{Eq:J(k,q)}-\ref{ep}) are $K$ and $q$. The derivatives entering in Eqs.~(\ref{meffNs}) can be expressed in terms of partial derivatives at constant $q$ or $K$
\begin{eqnarray}
\begin{array}{lll}
m_{eff}& = &
\left(\frac{\partial p}{\partial q}\right)_K^2
\left/
\left[
\left(\frac{\partial^2\varepsilon}{\partial q^2}\right)_K
-
V
\left(\frac{\partial^2 p}{\partial q^2}\right)_K
\right]
\right.
,\\
N_s&=&
- \left(\frac{\partial\varepsilon}{\partial \mu}\right)_q
+ \left(\frac{\partial \varepsilon}{\partial q}\right)_K
  \left(\frac{\partial V}{\partial \mu}\right)_q
\left/
\left(\frac{\partial V}{\partial q}\right)_K
\right.\\
V &=&
\left(\frac{\partial\varepsilon}{\partial q}\right)_K
\left/
\left(\frac{\partial p}{\partial q}\right)_K
\right.
\end{array}
\label{meffVNs:numerics}
.
\end{eqnarray}
First and second derivatives of $p$, $\varepsilon$ and $J(k,q)$ with respect to $q$ at fixed $K$ are found by solving additional integral equations which are obtained from Eqs.~(\ref{Eq:J(k,q)}-\ref{ep}) by differentiating with respect to the parameter $q$. Derivatives at fixed $q$ are calculated numerically.

\section{Results and discussion} 
\begin{figure}[tbp]
\includegraphics[width=\columnwidth,angle=0]{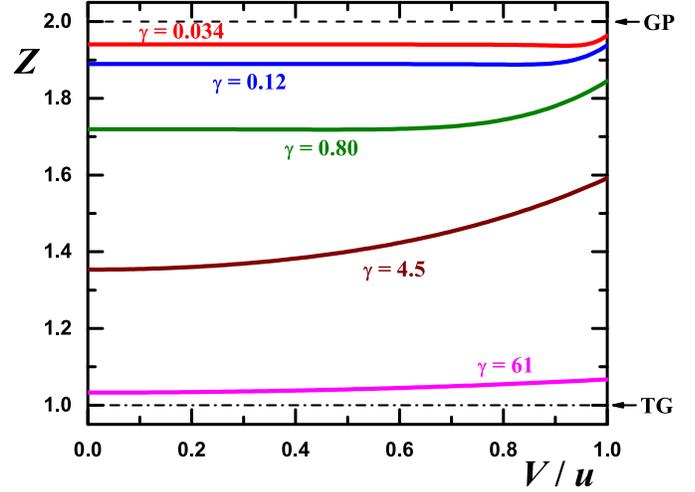}
\caption{(Color online) Parameter $Z=m_{eff}/(N_{s}m)$ as a function of velocity $V$ in units of the speed of sound $u$ at different values of interaction strength $\gamma$, from top to bottom, $\gamma = 0.034; 0.12; 0.80; 4.5; 61$ (corresponding to $K=20; 10; 3.3; 1; 0.1$). The dependence on $V$ disappears both in TG and GP limits.}
\label{Fig1}
\end{figure}

We calculated $N_s$ and $m_{eff}$ from Eqs.~(\ref{meffVNs:numerics}). As it was discussed above, the soliton dynamics completely described by the ratio $Z=m_{eff}/mN_{s}$. The dependency of $Z$ on velocity for different values of $\gamma$ presented in Fig.~\ref{Fig1}. The $V$-dependence disappears in the TG limit, where $Z=1$, and in GP limit, where $Z=2$ [see  equations (\ref{Eq:GP Ns}) and (\ref{Eq:TG meff})].

\begin{figure}[tbp]
\includegraphics[width=\columnwidth, angle=0]{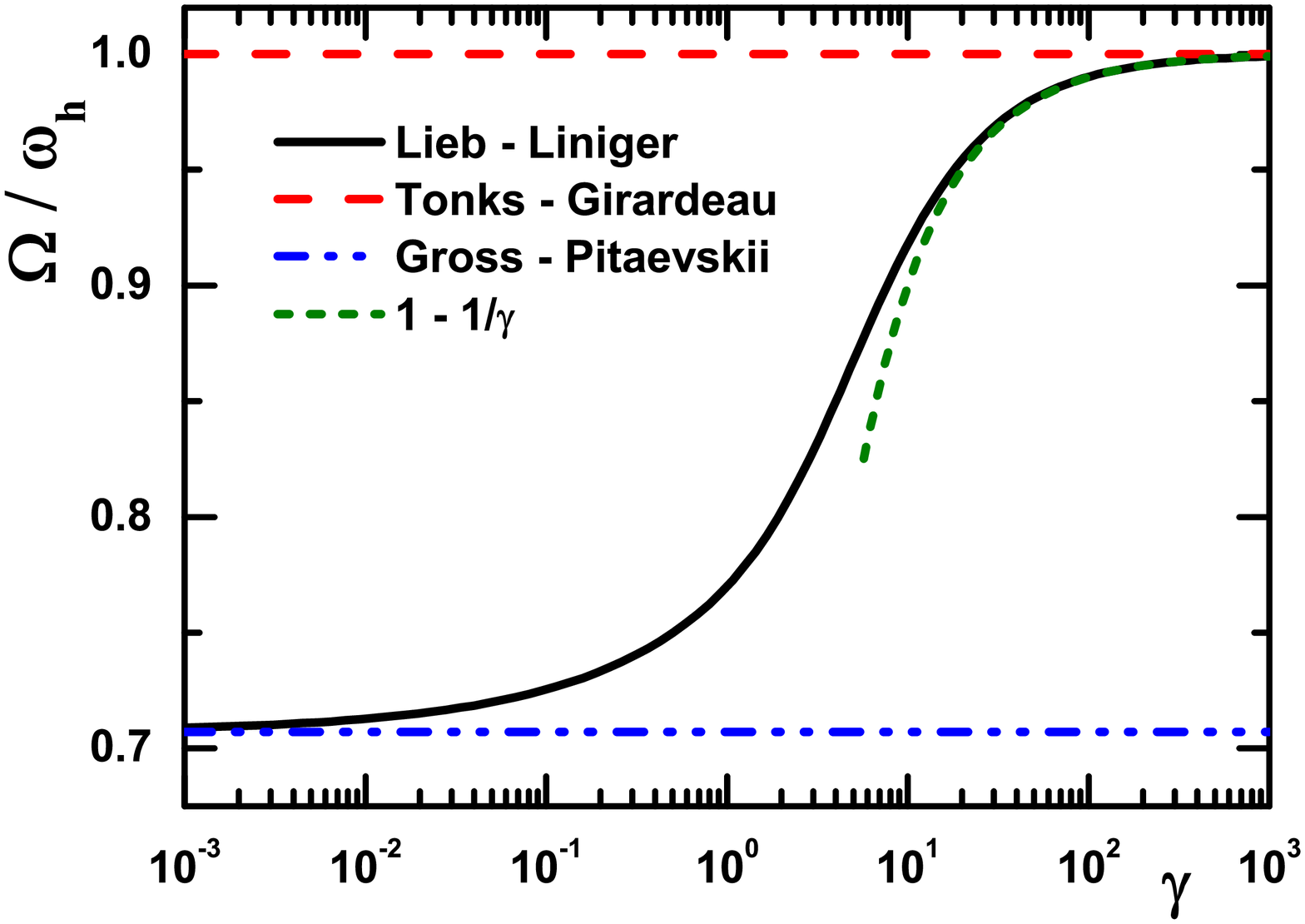}
\caption{(Color online) Solid line, frequency of oscillations $\Omega$ in units of the frequency $\omega_h$ of the harmonic oscillator as a function of the interaction parameter $\gamma$; dashed line, asymptotic value in TG limit; dash-dotted line, asymptotic value in GP limit; short-dashed line, perturbative solution of integral equations, $\Omega/\omega_h = 1-1/\gamma+...$.}
\label{Fig2}
\end{figure}

Probably the best way to experimentally verify our predictions is to measure the frequencies of oscillations $\Omega$ in a trap in different regimes. Figure~\ref{Fig1} shows the dependence of the frequency $\Omega $ of small oscillation on the interaction parameter $\gamma $. (The value of $\gamma$ should be taken for the center of the trap.)  One can see that the frequency continuously increases with increasing $\gamma $ from its GPE value $\omega_h/\sqrt{2}$ to the ideal Fermi gas value $\omega_h$. The most sharp change takes place at $\gamma \sim 3$. There are different ways of measuring the oscillations frequency. At moderately small values of $\gamma$, when a soliton still contains a large number of atoms, one can directly observe its motion, like in the experiments~\cite{Seng08,Lew99}. Instead at $\gamma\sim 1$ the number of atoms in a soliton is small, $|N_{s}|\sim 1$, one might exploit the confinement induced resonance (CIR)\cite{Olshanii98} in order to change the value of $\gamma $ in the course of an experiment. In typical one-dimensional experiments there is a number of elongated optical traps, created by standing waves. Initially solitons can be created by a phase imprinting method at small $\gamma $. Later the value of $\gamma $ might be increased by using CIR and soliton oscillations can be excited by a parametric modulation of the trap frequency. One expects to observe a resonance at the frequency of modulation $\Omega/2$. The resonance can be detected by heating of the gas.
\begin{figure}[tbp]
\includegraphics[width=\columnwidth, angle=0]{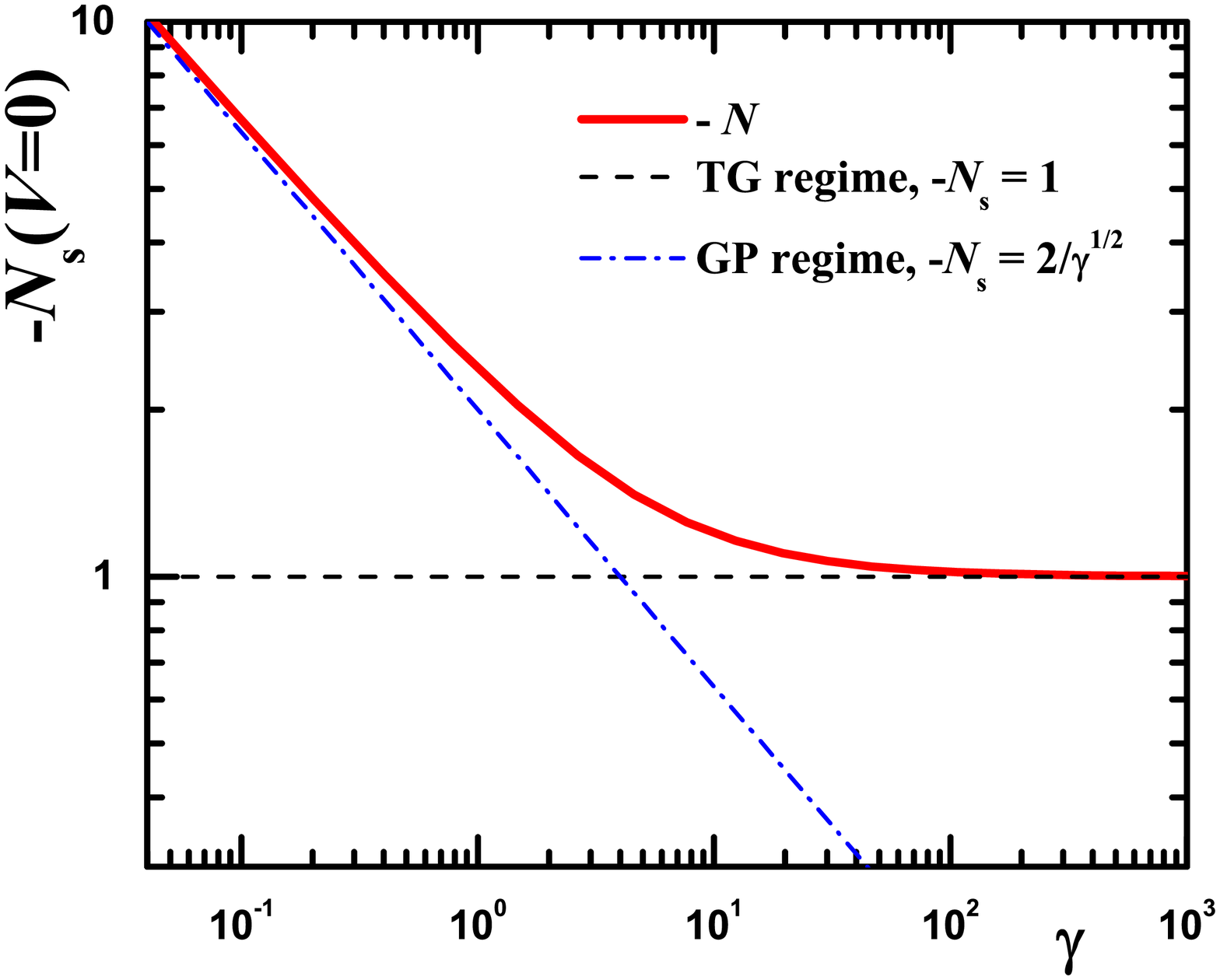}
\caption{(Color online) Solid line, number of particles $N_s$ in a stationary soliton ($V=0$) as a function of the interaction parameter $\gamma$; dashed line TG limit,  Eq.~(\ref{Eq:TG meff}); dash-dotted line GP limit, Eq.~(\ref{Eq:GP Ns}).}
\label{Fig3}
\end{figure}

\begin{figure}[tbp]
\includegraphics[width=\columnwidth, angle=0]{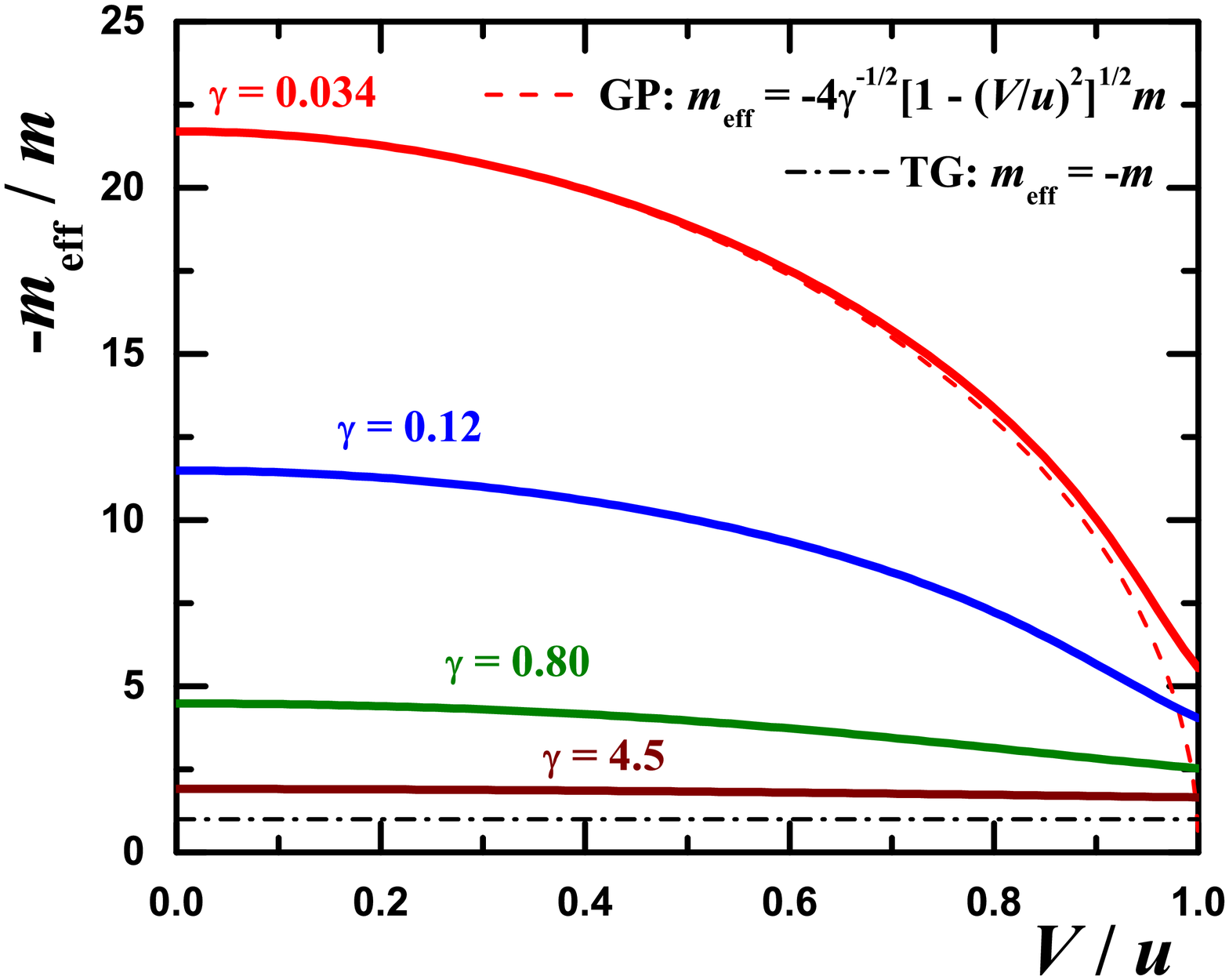}
\caption{(Color online) Solid lines, effective mass $m_{eff}$ in a as a function of velocity $V$ in units of speed of sound $u$ at different values of the interaction parameter $\gamma$; dashed line TG limit, Eq.~(\ref{Eq:TG meff}); dash-dotted line, GP limit, Eq.~(\ref{Eq:GP Ns}).}
\label{Fig4}
\end{figure}

\begin{figure}[tbp]
\includegraphics[width=\columnwidth, angle=0]{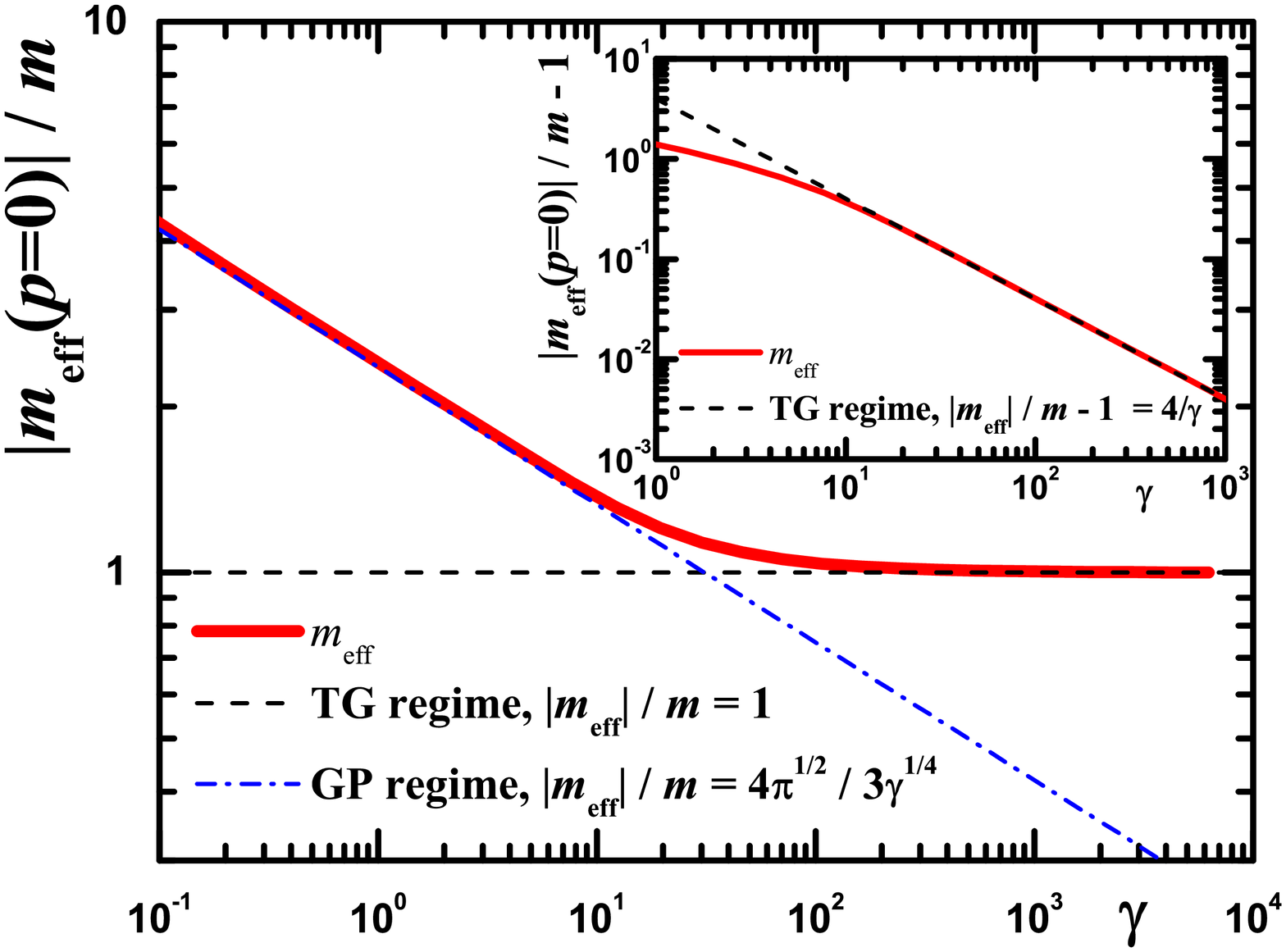}
\caption{(Color online) Effective mass of the ``fast'' soliton with $p=0$. Dashed line shows the asymptotic law Eq.~(\ref{Nsp0}). The inset shows $m_{eff}/m-1$ in comparison with large $\gamma$ result of~\cite{Brand05}.}
\label{Fig5}
\end{figure}

The frequency $\Omega$ of small oscillation is given by Eq.~(\ref{Omega}). It is a quantity of a large importance as it can be observed experimentally. However, $N_s(V=0)$ and $m_{eff}$ are interesting on their own. The dependence of the number of particles in the soliton at rest, $N_s(V=0)$, on the interaction strength is shown in Fig.~\ref{Fig3}. We find that for small $\gamma$ (GP regime) $|N_s| \gg 1$ and the soliton is a macroscopic object, however  $|N_s|$ becomes of the order of 1 already at $\gamma \sim 1$. In Fig.~\ref{Fig4} we presented $m_{eff}$ as a function of velocity $V$ at different values $\gamma$.
At small $\gamma$ both $N_S$ and $ m_{eff}$ are quite well described by the GP result, Eq.~(\ref{Eq:GP Ns}). However, the situation is different for ``fast'' solitons with $p\rightarrow 0$, i.e. with $V\rightarrow u$. According to (\ref{Eq:GP Ns}) the effective mass of a soliton tends to zero, $m_{eff}\propto (u-V)^{1/2}$, correspondingly to the small amplitude of the soliton. However the calculations show that $m_{eff}$ tends to a finite value at $V\rightarrow u$. This means that the dispersion law of the soliton should has the expansion at $p\rightarrow 0$
\begin{equation}
\varepsilon (p)\approx up+\frac{p^2}{2|m_{eff}(p=0)|} \;.  \label{mstar}
\end{equation}
This relation is quite non-trivial, because presence of the $p^2$ contradicts the GPE. Indeed, according to GPE $\varepsilon -up\propto p^{5/3}$. The existence of the $p^2$ term in the spectrum of 1D bosons was established by Imambekov, Schmidt, and Glazman (see \cite{GRMP}, Eq.~(50)). Such a term exists both for upper and lower branches of elementary excitations. The effective mass is the same in the absolute value for two branches. (See \cite{GRMP}, a paragraph after Eq. (172).) Simple calculation permits to present the result of \cite{GRMP} as
\begin{equation}
\left\vert m_{eff}(p=0)\right\vert^{-1}=\frac{3}{4}\sqrt{\frac{u}{\pi \hbar m\rho }}\left( 1+\frac{\rho^2}{3u^2}\frac{d(u^2/\rho )}{d\rho} \right) \;.
\end{equation}
In the GPE regime $\gamma \ll 1$, velocity of sound $u\propto \rho^{1/2}$, and the second term disappears. Then
\begin{equation}
|m_{eff}(p=0)|=\frac{4\sqrt{\pi}}{3}\gamma^{-1/4} = 2.36\gamma^{-1/4}\;
\label{Nsp0}
\end{equation}
(see \cite{GRMP}, the paragraph next to Eq. (172)). In Fig.~\ref{Fig5} we test the obtained result by showing the dependence of $|m_{eff}(p=0)|\gamma^{1/4}$ on $\gamma$. One can see a good agreement for the coefficient in Eq.~(\ref{Nsp0}) in GP limit. It is possible to show that the presence of the $p^2$ term in dispersion does not violate the GPE relation $m_{eff}=2mN_{s} $. Thus, this peculiar effect has no influence on the equation of motion~(\ref{Newt}).

In the inset of Fig.~\ref{Fig5} we test the expansion of $m_{eff}$ in TG limit. To do so we plot the quantity $|m_{eff}|/m-1$ on a log-log scale and compared it with  the expression, obtained for $\gamma \gg 1$ in \cite{Brand05} in the Hartree-Fock approximation.

To conclude, by using the exact Lieb-Liniger theory we investigated physical characteristics of Lieb II soliton-like branch of excitations. The frequency of oscillations, effective mass and number of atoms in the soliton are calculated. Direct numerical calculations confirmed violation applicability of the GP equation at small momentum $p$ in accordance with exact theory. The experimental possibility of the verification of the calculations is discussed.

Authors thank J.~Brand, L.D.~Faddeev, D.M.~Gangardt, and L.I.~Glazman for fruitful discussions. G.E.A. acknowledges support from the Spanish MEC through the Ramon y Cajal fellowship program. L.P.P. acknowledges support by ERC through the QGBE grant and by the Italian MIUR through the PRIN-2009 grant.


\begin{thebibliography}{99}
\bibitem{Essl03}
  \Name{Moritz H., St\"{o}ferle T., K\"{o}hl M. \and Esslinger T.}
  \REVIEW{Phys. Rev. Lett.}{91}{2003}{250402}.

\bibitem{Esteve06}
  \Name{Esteve J., Trebbia J.-B., Schumm T., Aspect A., Westbrook C. I. \and Bouchoule I.}
  \REVIEW{Phys. Rev. Lett.}{96}{2006}{130403}.

\bibitem{Gir60}
  \Name{Girardeau M.}
  \REVIEW{J. Math. Phys. (NY)}{1}{1960}{516}.

\bibitem{Tonks36}
  \Name{Tonks L.}
  \REVIEW{Phys. Rev.}{50}{1936}{955}.

\bibitem{LiebLiniger63}
  \Name{Lieb E. H. \and Liniger W.}
  \REVIEW{Phys. Rev.}{130}{1963}{1605}.

\bibitem{Lieb63}
  \Name{Lieb E. H.}
  \REVIEW{Phys. Rev.}{130}{1963}{1616}.

\bibitem{Fadd76}
  \Name{Kulish P. P., Manakov S. V. \and Faddeev L. D.}
  \REVIEW{Theoreticheskaya i Matematicheskaya Fizika}{28}{1976}{38};
  [\REVIEW{Theoretical and Mathematical Physics}{28}{1976}{615}].

\bibitem{Ish80}
  \Name{Ishikawa M. \and Takayama H.}
  \REVIEW{J. Phys. Soc. Jap.}{49}{1980}{1242}.

\bibitem{Tsu71}
  \Name{Tsuzuki T.}
  \REVIEW{J. Low Temp. Phys.}{4}{1971}{441}.

\bibitem{Sato12}
 \Name{Sato J., Kanamoto R., Kaminishi E. and Deguchi T.}
  \REVIEW{arXiv:1204.3960v1}{}{2012}{}.

\bibitem{Carr10}
\Name{Kanamoto R., Carr L.D. and  Ueda M.}
 \REVIEW{Phys. Rev. A}{81}{2010}{023625}.

\bibitem{Shl99}
  \Name{Fedichev P. O., Muryshev A. E. \and Shlyapnikov G. V.}
  \REVIEW{Phys. Rev. A}{60}{1999}{3220}.

\bibitem{KonPit04}
  \Name{Konotop V. V. \and Pitaevskii L.}
  \REVIEW{Phys. Rev. Lett.}{93}{2004}{240403}.

\bibitem{Scott11}
  \Name{Scott R. G., Dalfovo F., Pitaevskii L. P. \and Stringari S.}
  \REVIEW{Phys. Rev. Lett.}{106}{2011}{185301}.

\bibitem{Schecter12}
  \Name{Schecter M., Gangardt D. M. \and Kamenev  R. G.}
  \REVIEW{Ann. Phys.}{327}{2012}{639}.

\bibitem{Shevch}
  \Name{Shevchenko S.}
  \REVIEW{Sov. J. Low Temp. Phys.}{14}{1988}{553}.

\bibitem{Ang00}
  \Name{Busch Th. \and Anglin J. R.}
  \REVIEW{Phys. Rev. Lett.}{84}{2000}{2298}.

\bibitem{Seng08}
  \Name{Becker C., Stellmer S., Soltan-Panani P., D\"{o}rscher S., Baumert M., Richter E.-M., Kronj\"{a}ger J., Bongs K. \and Sengstock K.}
  \REVIEW{Nature Phys.}{4}{2008}{496}.

\bibitem{Gang12}
  \Name{Wadkin-Snaith D. C. \and Gangardt D. M.}
  \REVIEW{Phys. Rev. Lett.}{108}{2012}{085301}.

\bibitem{Lew99}
  \Name{Burger S., Bongs K., Dettmer S., Ertmer W., Sengstock K., Sanpera A., Shlyapnikov G. V. \and Lewenstein M.}
  \REVIEW{Phys. Rev. Lett.}{83}{1999}{3577}.

\bibitem{Olshanii98}
  \Name{Olshanii M.}
  \REVIEW{Phys. Rev. Lett.}{81}{1998}{938}.

\bibitem{GRMP}
  \Name{Imambekov A., Schmidt T. L. \and Glazman L. I.}
  \REVIEW{Rev. Mod. Phys.}{84}{2012}{1253}.

\bibitem{Brand05}
  \Name{Brand J. \and Cherny A.}
  \REVIEW{Phys. Rev. A}{72}{2005}{033619}.

\end{thebibliography}
\end{document}